\documentclass[english,aps,twocolumn]{revtex4}
\usepackage[T1]{fontenc}
\usepackage[latin9]{inputenc}
\usepackage{float}
\usepackage{graphicx}
\usepackage{amssymb}

\usepackage{babel}

\begin{document}

\title{Vibrational wave packet induced oscillations in two-dimensional electronic
spectra. \\ II. Theory}

\author{Tom\'{a}\v{s} Man\v{c}al$^{1}$, Alexandra Nemeth$^{2}$, Franz Milota$^{3}$, Vladim\'{i}r Luke\v{s}$^{4}$, Harald F. Kauffmann$^{5}$ and Jaroslaw Sperling$^{6}$}

\affiliation{$^{1}$Faculty of Mathematics and Physics, Charles University in
Prague, Ke Karlovu 5, Prague 121 16, Czech Republic}

\affiliation{$^{2}$Faculty of Chemistry, University of Vienna, W�hringerstrasse
42, 1090 Vienna, Austria}

\affiliation{$^{3}$Lehrstul f\"{u}r BioMolekulare Optik, Ludwig-Maxmilians-Universit\"{a}t, Oettingenstrasse 67, 80538 M\"{u}nchen, Germany}

\affiliation{$^{4}$Department of Chemical Physics, Slovak Technical University,
Radlinsk�ho 9, 81237 Bratislava, Slovakia}

\affiliation{$^{5}$Ultrafast Dynamics Group, Faculty of Physics, Vienna University of Technology, Wiedner Hauptstrasse 8-10, 1040 Vienna, Austria}

\affiliation{$^{6}$Newport Spectra-Physics, Guerickeweg 7, 64291 Darmstadt, Germany}

\begin{abstract}
We present a theory of vibrational modulation of two-dimensional coherent
Fourier transformed electronic spectra. Based on an expansion of the
system's energy gap correlation function in terms of Huang-Rhys factors,
we explain the time-dependent oscillatory behavior of the absorptive
and dispersive parts of two-dimensional spectra of a two-level electronic
system, weakly coupled to intramolecular vibrational modes. The theory
predicts oscillations in the relative amplitudes of the rephasing
and non-rephasing parts of the two-dimensional spectra, and enables
to analyze time dependent two-dimensional spectra in terms of simple
elementary components whose line-shapes are dictated by the interaction
of the system with the solvent only. The theory is applicable to both
low and high energy (with respect to solvent induced line broadening)
vibrations. The results of this paper enable to qualitatively explain
experimental observations on low energy vibrations presented in the
preceding paper {[}A. Nemeth et al, arXiv:1003.4174v1{]}
and to predict the time evolution of two-dimensional spectra in ultrafast
ultrabroad band experiments on systems with high energy vibrations.
\end{abstract}
\maketitle

\section{Introduction}

Multi-dimensional ultrafast spectroscopic techniques open an increasingly
broad experimental window into the dynamics of molecular systems on
femtosecond and picosecond time scales. Important experimental insights
have been obtained by studying dynamics of small systems such as vibrations of small molecules, peptides and proteins \cite{HochstrasserReview09}, hydrogen
bonding networks in water \cite{Fecko06a} and peptides \cite{Hamm06}, and
electronic transitions in dye molecules \cite{Brixner04b,Nemeth08a}. At the same time, multi-dimensional spectroscopy have proven its strength on complex systems ranging from conjugated polymers \cite{Scholes09a,Scholes09b},
nanotubes \cite{Nemeth09b}, protein chromophore complexes \cite{Brixner05a,Zigmantas06a,Engel07a,Lee07},
all the way to solid state systems \cite{Nelson09}. The experimental work has been closely interlinked with theoretical efforts \cite{JonasReview,Cho05a,Mancal06a,Pisliakov06a,Cho08a,Cho05b,Brugemann06,Brugemann07,Abramavicius09a,Abramavicius08a,YCC07a,YCC08a,Egorova07,Egorova08}
which provided interpretation to this rich source of experimental
information. One of the most interesting recent topics in this field
was stimulated by the observation of oscillatory modulations of two-dimensional
(2D) spectra, that were assigned to the presence and time-evolution
of electronic wavepackets \cite{Engel07a,Scholes99a,Pisliakov06a}.
While the oscillatory effects were predicted theoretically from exciton
theory \cite{Pisliakov06a,Brugemann06}, the details of the observed
processes suggest that theories traditionally applied to model energy
relaxation and transfer in these systems underestimate the role of
coherence effects. For example, the decoupling of exciton population
and coherence dynamics (secular approximation) seems not to be a valid
approximation for the photosynthetic protein FMO \cite{Engel07a}.

Favorable properties of the studied multichromophoric systems, together
with a partial success of exciton theory allowed to exclude vibrational
coherence as a source of the oscillatory effects in the experiments
on FMO \cite{Engel07a}. In general however, a complementary understanding
of the time-dependent spectral features of vibrational origin are
clearly required in order to be able to disentangle the two effects.
To this end, the topic of this and the preceding experimental paper,
Ref. \cite{Nemeth09a} (paper I), is a system which can exhibit only
vibrational coherence, a two-level electronic system interacting with
well-defined intramolecular vibrational modes.

For the description of non-linear spectroscopic signals of simple few electronic level systems with negligible relaxation between the levels, the non-linear
response functions derived in the second order cumulant approximation
(see e.g. \cite{MukamelBook}) provide a framework of unprecedented
utility. The non-linear optical response of such a
system can be expressed in terms of line-broadening functions, which
are related to the electronic energy gap correlation function \cite{MukamelBook}.
Few electronic level systems provide models which enable to disentangle
the content of experimental data otherwise difficult to interpret. For example,
three pulse photon echo peakshift was identified by Cho et al. \cite{Cho96}
to reveal the energy gap correlation function. Two-color photon echo
peakshift has been found to measure excitonic coupling in a dimer
\cite{Yang99} or differences between energy gap correlation functions
of two chromophores \cite{Mancal04a}. Basic 2D spectral line-shapes
were classified by Tokmakoff \cite{Tokmakoff2000} based on a two-level
system, and finally, the relation between the non-linear spectroscopic signals
and the normalized energy gap correlation function was studied by
de Boeij et al. \cite{deBoeij96} and for 2D spectra by Lazonder
et al. \cite{Pshenichnikov06}, to name but a few examples.

In some of these works, insightful information was obtained by expanding
the line-shape functions (or energy gap correlation functions) in
terms of a suitable parameter, e.g. the time delay between two interactions.
In the present study, we follow the same general approach and expand
the line-shape correlation functions (describing the interaction of
an electronic transition with an intra-molecular vibrational mode)
in terms of another suitable parameter, the Huang-Rhys factor $\lambda/\omega$.
Here, $\lambda$ is the reorganization energy and $\omega$ the frequency
of the mode. The applicability of a low order expansion of the response
functions in terms of the Huang-Rhys factor to the description of
the 2D experiment is obviously system-dependent but for the dye molecule
studied in Refs. \cite{Nemeth09a} and \cite{Nemeth08a} the qualitative
results of the theory presented in this paper were confirmed by a
full response function simulation of the 2D spectrum \cite{Nemeth08a}.

One of the first theoretical analyses of the vibrational line-shapes in electronic 2D spectra was done by Egorova et al. \cite{Egorova07,Egorova08} using the density matrix propagation method. We complement this work by performing a detailed analysis of a two-level electronic system within the response function theory, where vibrational levels are not considered explicitely. Unlike in Refs. \cite{Egorova07} and \cite{Egorova08} we consider both rephasing and non-rephasing parts of the 2D spectrum, and we study time dependent modulations of the 2D spectrum by slow vibrational modes.

This paper is organized as follows. In the next section, we shortly
review two-dimensional Fourier transformed spectroscopy and its theoretical
description. In Section \ref{sec:Third-order-Non-linear} we introduce
the non-linear response functions and the line-shape functions which
form the basis of our description of the 2D spectra. In Section \ref{sec:Modulation-of-non-linear}
we decompose the total 2D line-shape into several components according
to their time dependence and derive elementary 2D line-shapes, which
can be used to construct the total 2D spectrum. The time evolution
of 2D spectra modulated by low and high energy modes are analyzed
in Section \ref{sec:Analysis-of-time}. Our conclusions are summarized
in Section \ref{sec:Conclusions}.

\section{Two-Dimensional Fourier Transformed Spectroscopy\label{sec:Two-Dimensional-Fourier-Transformed}}

The principles of Two-Dimensional Fourier Transformed Spectroscopy
(2D-FT) where described in detail in many references (see e.g. \cite{Brixner04b,JonasReview})
including the preceding paper, Ref. \cite{Nemeth09a}. Here, we sum
up results important for the theoretical modeling of the 2D-FT spectra
in this paper.

In 2D-FT we use three laser pulses with wave-vectors ${\bf k}_{1}$,
${\bf k}_{2}$, and ${\bf k}_{3}$ arriving at the sample at times
$\tau_{1}$, $\tau_{2}$, and $\tau_{3}$, respectively. In case of
sufficiently weak excitation pulses, the intensity of the signal observed
in the direction of $-{\bf k}_{1}+{\bf k}_{2}+{\bf k}_{3}$ is of
third order in the total intensity of the incident light and the signal
can be well described by third order time-dependent perturbation theory.
The propagation of light through a material sample is influenced by
the light-induced polarization, which is proportional to a material
response function. The total third order response of the system to
an excitation by three successive interactions with an external electric
field of a laser is
$$
{\cal R}^{(3)}(t_{3},t_{2},t_{1})=\Theta(t_{3})\Theta(t_{2})\Theta(t_{1})
$$
\begin{equation}
\times \left(\frac{i}{\hbar}\right)^{3}\sum_{n=1}^{4}\left\{ R_{n}(t_{3},t_{2},t_{1})-R_{n}^{*}(t_{3},t_{2},t_{1})\right\} ,\label{eq:S3}
\end{equation}
where $\Theta(t)$ is the Heaviside step-function. The $R_{n}$ functions
are defined according to the standard notation of Ref. \cite{MukamelBook},
and represent different perturbative contributions (know as Liouville pathways) to the total response. The response is a function of three time
variables, where $t_{1}$ denotes the delay between the first and
second interaction, $t_{2}$ denotes the delay between the second
and third interaction, and $t_{3}$ is the delay between the last
interaction and the measurement. The polarization of the sample at
time $t_{3}$ after the arrival of the third pulse is given by a three-fold
convolution of the response function, Eq. (\ref{eq:S3}), with the
incoming electric fields. For very short pulses (impulsive limit)
one can identify times $t_{1}$ and $t_{2}$ with the delays between
the pulses ($t_{1}=|\tau_{2}-\tau_{1}|$, $t_{2}=|\tau_{3}-\tau_{2}|$)
and one can approximate the polarization by those parts of the above
expression that survive the so-called rotating wave approximation
(RWA) (see e.g. Ref. \cite{Brixner04b}). This yields $P_{s}^{(3)}(t_{3},t_{2},t_{1})\approx{\cal R}_{rwa}^{(3)}(t_{3},t_{2},t_{1})$.
With different ordering of the pulses ${\bf k}_{1}$ and ${\bf k}_{2}$,
${\cal R}_{rwa}^{(3)}$ contains different Liouville pathways. In
particular
\begin{equation}
P_{s}^{(3)}(t_{3},t_{2},t_{1})\approx-iR_{2}(t_{3},t_{2},t_{1})-iR_{3}(t_{3},t_{2},t_{1}),\label{eq:Reph}
\end{equation}
for $\tau_{1}<\tau_{2}$ and
\begin{equation}
P_{s}^{(3)}(t_{3},t_{2},t_{1})\approx-iR_{1}(t_{3},t_{2},t_{1})-iR_{4}(t_{3},t_{2},t_{1}),\label{eq:nonReph}
\end{equation}
for $\tau_{1}>\tau_{2}$.
As pointed out in the Introduction, we are interested in two level electronic systems, for which the number of unique pathways is limited to four \cite{MukamelBook}.
The electric field of the measured signal is proportional to the
polarization as
\begin{equation}
E_{s}^{(3)}(t_{3},t_{2},t_{1})\approx\frac{i\omega}{n(\omega)}P_{s}^{(3)}(t_{3},t_{2},t_{1}).\label{eq:Es}
\end{equation}
 In 2D-FT experiments, the signal travels into a direction different
from all incoming pulses, where it can be detected by a heterodyne
detection scheme without interfering radiation from the three excitation
pulses. The radiative prefactor $\omega/n(\omega)$ is removed and
the 2D spectrum is obtained by a double Fourier transform (experimentally
only one of the Fourier transforms is performed numerically, the other
is implicitly involved in the frequency resolved detection scheme)
\cite{JonasReview,Brixner04b}. Signals from both possible time orderings
of the ${\bf k}_{1}$ and ${\bf k}_{2}$ pulses (i.e. $\tau_{1}<\tau_{2}$
and $\tau_{1}>\tau_{2}$) are summed to form the 2D-FT spectrum yielding
\[
S(\omega_{3},t_{2},\omega_{1}) \]
\[
=FT_{2D}^{(-)}[R_{2}(t_{3},t_{2},t_{1})+R_{3}(t_{3},t_{2},t_{1})](\omega_{3},\omega_{1})\]
\begin{equation}
+FT_{2D}^{(+)}[R_{1}(t_{3},t_{2},t_{1})+R_{4}(t_{3},t_{2},t_{1})](\omega_{3},\omega_{1}),\label{eq:2D_spect}\end{equation}
 where
\[
FT_{2D}^{(-)}[f(t_{3},t_{1})](\omega_{3},\omega_{1})
\]
\begin{equation}
=\int\limits _{0}^{\infty}dt_{3}\int\limits _{0}^{\infty}dt_{1}f(t_{3},t_{1})e^{i\omega_{3}t_{3}-i\omega_{1}t_{1}}\label{eq:FTminus}\end{equation}
\[
FT_{2D}^{(+)}[f(t_{3},t_{1})](\omega_{3},\omega_{1})
\]
 \begin{equation}
=\int\limits _{0}^{\infty}dt_{3}\int\limits _{0}^{\infty}dt_{1}f(t_{3},t_{1})e^{i\omega_{3}t_{3}+i\omega_{1}t_{1}}.\label{eq:FTplus}\end{equation}
The contributions stemming from different orderings of the pulses
posses different phase factors with respect to times $t_{1}$ and
$t_{3}$. Pathways $R_{2}$ and $R_{3}$ have a phase factor which
turns equal to zero for all frequencies at $t_{1}=t_{3}$. They are
referred to as rephasing pathways. Pathways $R_{1}$
and $R_{4}$ do not rephase in this sense and are referred to as non-rephasing.
Thus, in the impulsive limit the 2D-FT spectrum can be calculated
directly by applying a double Fourier transform to the rephasing and
non-rephasing components of the third order response function.

\section{Third order Non-linear response function\label{sec:Third-order-Non-linear}}

For a two-level electronic system, the third order response functions
can be derived in terms of a single quantity called line-shape function,
$g(t)$. The line-shape function is defined as

\begin{equation}
g(t)=\int\limits _{0}^{t}d\tau\int\limits _{0}^{\tau}d\tau'C(\tau'),\label{eq:goft}\end{equation}
 where $C(t)$ is the so-called energy gap (or frequency-frequency)
correlation function, describing the fluctuations of the electronic
transition due to its interaction with the degrees of freedom (DOF)
of the surrounding. Often it is advantageous to use another
function, usually denoted as $M(t)$, for the discussion of the influence of
the bath DOF onto the spectrum (see e.g. \cite{deBoeij96,Pshenichnikov06}). The $M(t)$ function is (for high temperatures) equal to the real part of the correlation function $C(t)$ normalized to $1$ at $t=0$.
An analysis of nonlinear spectra based on $M(t)$ was done also in the preceding paper, Ref. \cite{Nemeth09a}.
The response functions read in detail
\begin{equation}
R_{n}(t_{3},t_{2},t_{1})=|d|^{4}e^{i\omega_{eg}(t_{1}-t_{3})+f_{n}(t_{3},t_{2},t_{1})},\; n=2,3\label{eq:Rreph}\end{equation}
 \begin{equation}
R_{n}(t_{3},t_{2},t_{1})=|d|^{4}e^{-i\omega_{eg}(t_{1}+t_{3})+f_{n}(t_{3},t_{2},t_{1})},\; n=1,4,\label{eq:Rnreph}\end{equation}
where $d$ is the transition dipole moment between the electronic
ground and excited state, and we defined
\[
f_{1}(t_{3},t_{2},t_{1})=-g(t_{1})-g^{*}(t_{3})-g^{*}(t_{2})+g^{*}(t_{2}+t_{3})
\]
\begin{equation}
+g(t_{1}+t_{2})-g(t_{1}+t_{2}+t_{3}),\label{eq:R1}\end{equation}

\[
f_{2}(t_{3},t_{2},t_{1})=-g^{*}(t_{1})-g^{*}(t_{3})+g(t_{2})-g(t_{2}+t_{3}) \]
 \begin{equation}
 -g^{*}(t_{1}+t_{2})+g^{*}(t_{1}+t_{2}+t_{3}),\label{eq:R2}
 \end{equation}
 
\[
f_{3}(t_{3},t_{2},t_{1})=-g^{*}(t_{1})-g(t_{3})+g^{*}(t_{2})-g^{*}(t_{2}+t_{3}) \]
\begin{equation}
-g^{*}(t_{1}+t_{2})+g^{*}(t_{1}+t_{2}+t_{3}),\label{eq:R3}\end{equation}
 
\[
f_{4}(t_{3},t_{2},t_{1})=-g(t_{1})-g(t_{3})-g(t_{2})+g(t_{2}+t_{3}) \]
\begin{equation}+g(t_{1}+t_{2})-g(t_{1}+t_{2}+t_{3}).\label{eq:R4}\end{equation}
 Using Eqs. (\ref{eq:Reph}) to (\ref{eq:R4}) we can calculate the
impulsive response of a two-level electronic system interacting with
a rather general environment that is described by some correlation
function $C(t)$.

One of the most common models for the correlation function $C(t)$
is the so called Brownian oscillator model, which enables to describe
both overdamped solvent modes as well as underdamped intramolecular
vibrational modes. The Brownian oscillator model and its application
to calculation of optical spectra is extensively described in Ref.
\cite{MukamelBook}. In our considerations, we represent certain
interesting intramolecular modes by the underdamped version of this
model, assuming further that the damping of the oscillator is negligible.
For the description of the experimental results, Ref. \cite{Nemeth09a},
it seems to be a good approximation for times less than $1$~ps. The
corresponding energy gap correlation function thus reads \cite{MukamelBook}

\begin{equation}
C(t)=\lambda\omega\Xi(T)\cos\omega t-i\lambda\omega\sin\omega t,\label{eq:corrfce}\end{equation}
where $\Xi(T)=\coth\left(\frac{\hbar\omega}{2k_{B}T}\right)$. From Eq. (\ref{eq:corrfce}) it follows that
\begin{equation}
g(t)=i\lambda t+i\frac{\lambda}{\omega}\sin\omega t+\frac{\lambda}{\omega}\Xi(T)\left[1-\cos\omega t\right].\label{eq:osc_goft}\end{equation}
Here, $\lambda$ is the reorganization energy and $\omega$ is the
frequency of the mode, respectively. The overdamped Brownian oscillator
model is used for modeling the solvent correlation function. We will
be interested in the properties of the corresponding solvent $g(t)$-function
for $t>\tau_{c}$ only, where $\tau_{c}$ is some characteristic correlation
time of the solvent. Particularly, if the solvent energy gap correlation
function does not have any slow static component, i.e. if so-called
static disorder of the system is negligible, we can write for $t>\tau_{c}$
(see Ref. \cite{MukamelBook})
\begin{equation}
g(t)=g_{const}+\bar{g}(t),\label{eq:gt}
\end{equation}
where $\bar{g}(t)$ has the property
\begin{equation}
\bar{g}(t+\tau)=\bar{g}(t)+\bar{g}(\tau),\label{eq:gttau}
\end{equation}
i.e. for sufficiently long times, the function $g(t)$ is linear.
This property will be use below to simplify Eqs. (\ref{eq:R1}) to
(\ref{eq:R4}).

\section{Modulation of non-linear response by vibrational modes\label{sec:Modulation-of-non-linear}}

\subsection{Third order response of weakly coupled vibrational modes}

Let us consider an electronic two-level system that is coupled to
an environment consisting of one dominant underdamped vibrational
mode and a bath with a macroscopic number of DOF. The latter DOF are
represented by an overdamped Brownian oscillator model and its corresponding
line-shape function $g_{bath}(t)$ which has the property of Eq. (\ref{eq:gt}).
The total energy gap correlation function of the system will be assumed
in the form\begin{equation}
g(t)=g_{vib}(t)+g_{bath}(t),\label{eq:goft_split}\end{equation}
 where $g_{vib}(t)$ is defined by Eq. (\ref{eq:osc_goft}). The response
functions, Eqs. (\ref{eq:Rreph}) and (\ref{eq:Rnreph}), can thus
be factorized into pure solvent and pure vibrational parts. Since
we deal here with one electronic transition only, we assume $d=1$
for the rest of the paper. If the Huang-Rhys factor $\lambda/\omega$
(i.e. the coupling of the vibrational mode to the electronic transition)
is sufficiently small, we can expand the vibrational contribution
and write to first order in $\lambda/\omega$
\[R_{n}(t_{3},t_{2},t_{1}) \approx e^{i(\omega_{eg}+\lambda)(t_{1}-t_{3})}\bar{R}_{n}(t_{3},t_{2},t_{1})\]
\begin{equation}
\times\left[1+\frac{\lambda}{\omega}F_{n}(t_{1},t_{2},t_{3})\right],\; n=2,3\label{eq:R2_app}\end{equation}
\[
R_{n}(t_{3},t_{2},t_{1})\approx e^{-i(\omega_{eg}+\lambda)(t_{1}+t_{3})}\bar{R}_{n}(t_{3},t_{2},t_{1})\]
\begin{equation}
\times\left[1+\frac{\lambda}{\omega}F_{n}(t_{1},t_{2},t_{3})\right],\; n=1,4.\label{eq:R1_app}\end{equation}
Here, the $\bar{R}_{n}(t_{3},t_{2},t_{1})$ functions comprise the
pure solvent response (except of the optical frequency factor $e^{-i\omega_{eg}(t_{3}\pm t_{1})}$,
cf. Eq. (\ref{eq:R14_solv} and \ref{eq:R23_solv})), and
\[
F_{n}(t_{1},t_{2},t_{3})=K_{n}(t_{1},t_{3})+H_{n}(t_{1},t_{3})\cos\omega t_{2}
\]
\begin{equation}
+G_{n}(t_{1},t_{3})\sin\omega t_{2},\label{eq:Ffce}
\end{equation}
 with $K$, $H$, and $G$ functions discussed below. We used the fact
that if Eq. (\ref{eq:osc_goft}) is inserted into Eqs. (\ref{eq:R1})
to (\ref{eq:R4}), the first two terms we obtain are $f_{n}(t_{3},t_{2},t_{1})=i\lambda(t_{3}\pm t_{1})-2\lambda\Xi(T)/\omega+\dots$.
These two terms are kept in the exponential of Eqs. (\ref{eq:R2_app})
and (\ref{eq:R1_app}), and the rest is expanded in terms of $\lambda/\omega$.
The only formal differences between the rephasing and non-rephasing
response functions, Eqs. (\ref{eq:R2_app}) and (\ref{eq:R1_app}),
are the different phase factors in times $t_{1}$ and $t_{3}$.

The expansion of the exponential function in Eqs. (\ref{eq:R2_app}) and (\ref{eq:R1_app}), $\exp\{\frac{\lambda}{\omega}F_{n}\} \approx 1+\frac{\lambda}{\omega}F_{n}$, requires $\frac{\lambda}{\omega}F_{n}$ to be a small number. The function $F_{n}$ is an oscillating function bound between certain values which represent the points of the extreme deviation from the exact expression for the response functions. One of such values occurs e.g. at $t_{1}=0$ and $t_{3}=0$, where $F_{n} = 2\Xi(T)$. The validity of the approximation is therefore limited also by temperature. For high temperatures, however, i.e. if $\hbar\omega \gg 2k_{B}T$ and correspondingly $\Xi(T)\approx 1$,
the validity is limited only by the value of the Huang-Rhys factors. The 2D spectra that are the main subject of this paper are obtained by Fourier transforms in times $t_1$ and $t_3$. In applying the approximation it is therefore important that the approximated response function retains the correct periodicity in $t_1$ and $t_3$. The expansion can be generalized for several vibrational modes, but one has to bear in mind that the factor limiting the validity of the approximation might be the sum of Huang-Rhys factors of all expanded modes.

Using the well-known goniometric relations\begin{equation}
\sin(\alpha+\beta)=\cos\alpha\sin\beta+\sin\alpha\cos\beta,\label{eq:sine}\end{equation}
 \begin{equation}
\cos(\alpha+\beta)=\cos\alpha\cos\beta-\sin\alpha\sin\beta\label{eq:cosine}\end{equation}
 the $K$, $H$, and $G$ functions can be expanded into a linear
combination of simple $\cos$ and $\sin$ functions. The actual forms of these functions can be found in Ref. \cite{Nemeth08a} or in
Appendix \ref{sec:AppA}.
%
%
The $t_{2}-$dependence of the response related to vibrational modes
can thus be approximately separated into stationary, sine oscillating
and cosine oscillating parts, using relatively simple factors of Eqs.
(\ref{eq:K1}) to (\ref{eq:G4}) from Appendix \ref{sec:AppA}.

Due to the properties of the solvent line-shape function $g_{bath}(t)$,
namely Eqs. (\ref{eq:gt}) and (\ref{eq:gttau}), the response of the solvent is $t_{2}$-independent
for long times $t_{2}$. Thus, if $t_{2}>\tau_{c}$ we can write\begin{equation}
\bar{R}_{1}(t_{3},t_{2},t_{1})=\bar{R}_{4}(t_{3},t_{2},t_{1})=e^{-g_{bath}(t_{1})-g_{bath}(t_{3})},\label{eq:R14_solv}\end{equation}
 \begin{equation}
\bar{R}_{2}(t_{3},t_{2},t_{1})=\bar{R}_{3}(t_{3},t_{2},t_{1})=e^{-g_{bath}^{*}(t_{1})-g_{bath}(t_{3})}.\label{eq:R23_solv}\end{equation}
 This enables us to drop the $t_{2}$-dependence of the solvent response
function in the following discussion, bearing in mind that we always
assume $t_{2}>\tau_{c}$.

\subsection{Two-dimensional Fourier-transformed spectra}

The 2D-FT spectrum was defined in Section \ref{sec:Two-Dimensional-Fourier-Transformed}.
By the double Fourier-transforms, Eqs. (\ref{eq:FTminus}) and (\ref{eq:FTplus}),
we obtain the total spectrum and its rephasing and non-rephasing parts
at a given delay $t_{2}$. The expansions in Eqs. (\ref{eq:R2_app})
and (\ref{eq:R1_app}) suggest that the 2D spectrum is composed of
a vibration-free spectral shape originating from the solvent response
function $\bar{R}$ and some vibrational contributions. The pure solvent
contributions read
\[
Z_{n}^{solv}(\omega_{3},\omega_{1})=FT_{2D}^{(-)}[\bar{R}_{n}(t_{3},t_{1})
\]
\begin{equation}
\times e^{i\omega_{eg}(t_{1}-t_{3})}](\omega_{3},\omega_{1}),\quad n=2,3,\label{eq:Z0}\end{equation}

\[
Y_{n}^{solv}(\omega_{3},\omega_{1})=FT_{2D}^{(+)}[\bar{R}_{n}(t_{3},t_{1}) \]
 \begin{equation}
 \times e^{-i\omega_{eg}(t_{1}+t_{3})}](\omega_{3},\omega_{1}),\quad n=1,4.\label{eq:Y0}\end{equation}
 The solvent 2D-FT spectrum is therefore given by\[
S_{solv}(\omega_{3},\omega_{1})=Z_{2}^{solv}(\omega_{3},\omega_{1})+Z_{3}^{solv}(\omega_{3},\omega_{1})\]
 \begin{equation}
+Y_{1}^{solv}(\omega_{3},\omega_{1})+Y_{4}^{solv}(\omega_{3},\omega_{1}).\label{eq:2D_solvent}\end{equation}
 Considering Eqs. (\ref{eq:R14_solv}) and (\ref{eq:R23_solv}) we
see that one can write
\[
Z(\omega_{3},\omega_{1})\equiv Z_{2}^{solv}(\omega_{3},\omega_{1})
\]
\begin{equation}
=Z_{3}^{solv}(\omega_{3},\omega_{1})={\cal G}(\omega_{3}){\cal G}^{*}(\omega_{1}),\label{eq:Z0_byG}\end{equation}
 \[
 Y(\omega_{3},\omega_{1})\equiv Y_{1}^{solv}(\omega_{3},\omega_{1})
 \]
 \begin{equation}
=Y_{4}^{solv}(\omega_{3},\omega_{1})={\cal G}(\omega_{3}){\cal G}(\omega_{1}),\label{eq:Y0_byG}\end{equation}
 where we defined 
\begin{equation}
{\cal G}(\omega)=\int\limits _{0}^{\infty}dt\; e^{-g(t)+i(\omega-\omega_{eg})t}.\label{eq:defG}
\end{equation}
 Solvent contributions to the 2D spectrum can thus be expressed using
just two elementary 2D line-shapes $Z(\omega_{3},\omega_{1})$ and
$Y(\omega_{3},\omega_{1})$, which can be further simplified. The
real part of Eq. (\ref{eq:defG}) is directly proportional to the
absorption at frequency $\omega$. Consequently, one can conclude that the real part of the total 2D line-shape, $\mathrm{Re}(Z+Y) = \mathrm{Re}({\cal{G}}(\omega_3))\mathrm{Re}({\cal{G}}(\omega_1))$, represents an absorption--absorption correlation plot, while its imaginary part, $\mathrm{Im}(Z+Y) = \mathrm{Im}({\cal{G}}(\omega_3))\mathrm{Re}({\cal{G}}(\omega_1))$, corresponds to an absorption--refraction correlation plot. For simple symmetric line shapes with a maximum at some frequency $\omega_{max}$, e.g. for a Lorentzian, $Z$ and $Y$ are related by flipping of the line shape with respect to the $\omega_1$-axis at $\omega_{1}=\omega_{max}$.

Now we will use elementary rephasing and non-rephasing 2D spectral
line-shapes, Eqs. (\ref{eq:Z0_byG}) and (\ref{eq:Y0_byG}), to construct
a $t_{2}-$dependent line-shape of a molecule possessing a vibrational
mode with a sufficiently small Huang-Rhys factor. To this end, we
use the fact that the vibrational part of the response function consists
of the $t_{2}-$independent functions $K_{n}(t_{1},t_{3})$, $H_{n}(t_{1},t_{3})$
and $G_{n}(t_{1},t_{3})$ that contain only sines and cosines of various
combinations of arguments $\omega t_{1}$ and $\omega t_{3}$. The
whole $t_{2}-$dependence is contained in Eq. (\ref{eq:Ffce}). While
performing a Fourier transform of the response functions, Eqs. (\ref{eq:R2_app})
and (\ref{eq:R1_app}), we Fourier transform the solvent response
function multiplied by the cosine and sine terms from the functions
$K_{n}(t_{1},t_{3})$, $H_{n}(t_{1},t_{3})$ and $G_{n}(t_{1},t_{3})$.
The total rephasing spectrum can thus be constructed from six possible
vibrational contributions. We have e. g.
\[
Z_{cos}^{(3)}(\omega_{3},\omega_{1})=FT_{2D}^{(-)}[e^{i(\omega_{eg}+\lambda)(t_{1}-t_{3})}
\]
\begin{equation}
\times\bar{R}_{2}(t_{3},t_{1})\cos\omega t_{3}](\omega_{3},\omega_{1})\label{eq:Ztcos}.\end{equation}
The remaining expressions of the same kind are summarized in the first part of Appendix \ref{sec:AppB}.
From now on, we use only $\bar{R}_{2}(t_{3},t_{1})$, because $\bar{R}_{2}(t_{3},t_{1})=\bar{R}_{3}(t_{3},t_{1})$
for $t_{2}>\tau_{c}$ (cf. Eq. (\ref{eq:R23_solv})), and we drop
the lower index for the line-shapes $Z$ and $Y$. Similarly we obtain
shapes $Y_{cos}^{(3)}$, $Y_{cos}^{(1)}$ etc. for the non-rephasing
part by applying $FT_{2D}^{(+)}$ to the $\bar{R}_{1}$ ($=\bar{R}_{4}$)
function.

The vibrational frequency $\omega$ is much smaller than the frequency
of the visible excitation light at which the 2D spectrum is measured
and consequently the sine and cosine in the Fourier transform
introduce only shifts of the spectral feature without significant
changes of its shape. Using the relations
\begin{equation}
\cos\omega t=\frac{e^{-i\omega t}+e^{i\omega t}}{2},\label{eq:cos2e}\end{equation}
 and \begin{equation}
\sin\omega t=i\frac{e^{-i\omega t}-e^{i\omega t}}{2},\label{eq:sin2e}\end{equation}
 we obtain e.g.\[
Z_{cos}^{(3)}(\omega_{3},\omega_{1})=\frac{1}{2}Z(\omega_{3}-\lambda-\omega,\omega_{1}-\lambda)\]
 \begin{equation}
+\frac{1}{2}Z(\omega_{3}-\lambda+\omega,\omega_{1}-\lambda),\label{eq:ZtcosInZ0}\end{equation}
 or\[
Z_{sin}^{(3)}(\omega_{3},\omega_{1})=\frac{i}{2}Z(\omega_{3}-\lambda-\omega,\omega_{1}-\lambda)\]
 \begin{equation}
-\frac{i}{2}Z(\omega_{3}-\lambda+\omega,\omega_{1}-\lambda),\label{eq:ZtsinInZ0}\end{equation}
 and analogously for other shapes. All shapes can thus be written
as simple linear combinations of the shifted basic shapes $Z$ and
$Y$. To save writing out of all arguments, we define\begin{equation}
\Omega_{3}\equiv\omega_{3}-\lambda,\quad\Omega_{1}\equiv\omega_{1}-\lambda,\label{eq:Omegas}\end{equation}
 and

\begin{equation}
Z^{(0+)}\equiv Z(\Omega_{3},\Omega_{1}+\omega),\label{eq:Zn0plus}\end{equation}
 \begin{equation}
Z^{(-+)}\equiv Z(\Omega_{3}-\omega,\Omega_{1}+\omega),\label{eq:Znplusminus}\end{equation}
 etc., where upper indices $+$, $-$ and $0$ of the newly defined
$Z$ functions indicate the additional shift by frequency $+\omega$,
$-\omega$ or $0\times\omega$, respectively. Using the above definitions
we can write all the contributions of the actual 2D spectrum e. g. as
\[
Z_{cos}^{(3)}(\Omega_{3},\Omega_{1})=\frac{1}{2}\left(Z^{(-0)}+Z^{(+0)}\right), \]
\begin{equation}
 Z_{sin}^{(3)}(\Omega_{3},\Omega_{1})=\frac{i}{2}\left(Z^{(-0)}-Z^{(+0)}\right)\label{eq:Zeq1}.\end{equation}
Again, the remaining equations are summarized in Appendix \ref{sec:AppB}.

According to Eqs. (\ref{eq:R2_app}) to (\ref{eq:Ffce}) (and assuming
$t_{2}>\tau_{c}$) the total response can be split into a $t_{2}-$independent
part (solvent and vibrational contribution given by functions $K_{n}(t_{3},t_{1})$
defined in Eqs. (\ref{eq:K1}) to (\ref{eq:K4})) and a part oscillating
with $t_{2}$ (given by functions $H_{n}(t_{3},t_{1})$ and $G_{n}(t_{3},t_{1})$
of Eqs. (\ref{eq:H1}) to (\ref{eq:G4})). Correspondingly, the total
2D-FT spectrum can then be split into a $t_{2}$-independent part,
$S^{0}$, and parts, $S^{cos}$, and $S^{sin}$, oscillating with
$\cos\omega t_{2}$ and $\sin\omega t_{2}$, respectively,
\[
S(\Omega_{3},t_{2},\Omega_{1})=S^{0}(\Omega_{3},\Omega_{1})+S^{cos}(\Omega_{3},\Omega_{1})\cos\omega t_{2}
\]
\begin{equation}
+S^{sin}(\Omega_{3},\Omega_{1})\sin\omega t_{2}.\label{eq:StotSpectrum}
\end{equation}
Collecting all parts of the response, we can express all contributions
to the total 2D-FT spectrum in terms of the $Z$ and $Y$ line-shapes
defined in Eqs. (\ref{eq:Z0_byG}) and (\ref{eq:Y0_byG}). We keep
the lower indices $R$ and $NR$ to denote the rephasing and non-rephasing
parts of the spectrum, respectively. The static rephasing part consisting
of the Fourier transform of the rephasing solvent response and functions
$K_{2}(t_{3},t_{1})$ and $K_{3}(t_{3},t_{1})$ reads
\begin{equation}
S_{R}^{0}(\Omega_{3},\Omega_{1})\approx \Big[Z+i\frac{\lambda}{\omega}Z_{sin}^{(1)}+\frac{\lambda}{\omega}\Xi(T)(Z_{cos}^{(1)}+Z_{cos}^{(3)})\Big].\label{eq:S0R}
\end{equation}
The sine oscillating rephasing part which is based on the $G_{2}(t_{3},t_{1})$
and $G_{3}(t_{3},t_{1})$ functions reads
\[
S_{R}^{sin}(\Omega_{3},\Omega_{1})\approx -\frac{\lambda}{\omega}\Big[Z_{sin}^{(3)}+Z_{sin}^{(1)}-Z_{sin}^{(3+1)}\Big]\]
 \begin{equation}
+i\frac{\lambda}{\omega}\Big[Z_{cos}^{(1)}-Z_{cos}^{(3+1)}\Big],\label{eq:SsinR}\end{equation}
 and the cosine oscillating rephasing part composed of the Fourier
transforms of the $H_{2}(t_{3},t_{1})$ and $H_{3}(t_{3},t_{1})$
functions can be written as
\[
S_{R}^{cos}(\Omega_{3},\Omega_{1})\approx \frac{\lambda}{\omega}\Big[Z_{cos}^{(3)}+Z_{cos}^{(1)}-Z_{cos}^{(3+1)}-Z\Big]\]
 \begin{equation}
+i\frac{\lambda}{\omega}\Big[Z_{sin}^{(1)}-Z_{sin}^{(3+1)}\Big].\label{eq:ScosR}\end{equation}
 The non-rephasing contributions are similarly based on the non-rephasing
functions with subindices $1$ and $4$ and read

\[
S_{NR}^{0}(\Omega_{3},\Omega_{1})\approx \Big[Y-i\frac{\lambda}{\omega}Y_{sin}^{(1)}
\]
\begin{equation}
+\frac{\lambda}{\omega}\Xi(T)(Y_{cos}^{(1)}+Y_{cos}^{(3)})\Big].\label{eq:S0NR}\end{equation}

\[
S_{NR}^{sin}(\Omega_{3},\Omega_{1})\approx \frac{\lambda}{\omega}\Xi(T)\Big[Y_{sin}^{(1)}+Y_{sin}^{(3)}-Y_{sin}^{(3+1)}\Big]\]
 \begin{equation}
+i\frac{\lambda}{\omega}\Big[Y_{cos}^{(1)}-Y_{cos}^{(3+1)}\Big],\label{eq:SsinNR}\end{equation}

\[
S_{NR}^{cos}(\Omega_{3},\Omega_{1})\approx -\frac{\lambda}{\omega}\Xi(T)\Big[Y_{cos}^{(3)}+Y_{cos}^{(1)}-Y_{cos}^{(3+1)}-Y\Big]\]
 \begin{equation}
+i\frac{\lambda}{\omega}\Big[Y_{sin}^{(1)}-Y_{sin}^{(3+1)}\Big].\label{eq:ScosNR}\end{equation}

The above equations enable to characterize completely the total time-dependent
2D-FT spectrum of a two-level electronic system weakly coupled to
a vibrational mode.

\section{Analysis of time-dependent 2D line-shapes\label{sec:Analysis-of-time}}

%
%
\begin{figure}[ht]
\includegraphics[clip,width=\columnwidth]{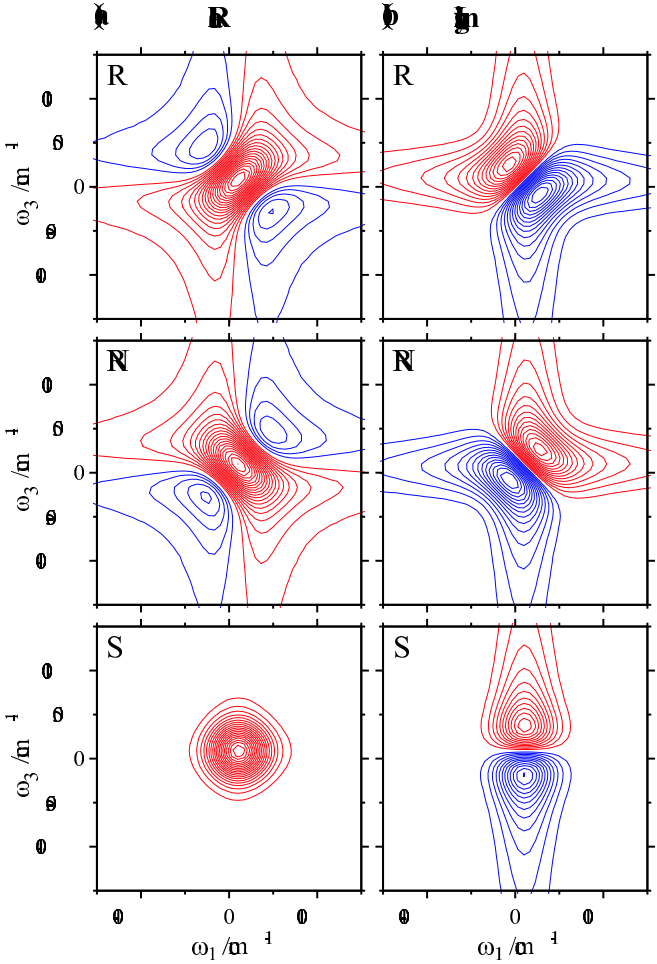}

\caption{\label{fig:shapeSlow}The stationary contributions (from top down)
$S_{R}^{0}(\omega_{3},\omega_{1})$, $S_{NR}^{0}(\omega_{3},\omega_{1})$
and their sum $S^{0}(\omega_{3},\omega_{1})$ for the modulation by
a low energy vibrational mode of $\omega=140$ cm$^{-1}$ and $\lambda=80$
cm$^{-1}$. The figures on the left present the real and the figures
on the right the imaginary parts of the spectra, respectively. Apart
from a slight asymmetry of the line-shapes with respect to the diagonal
or anti-diagonal line, these figures represent very well the pure
solvent line-shapes, too.}

\end{figure}

%
\begin{figure}[ht]
\includegraphics[clip,width=\columnwidth]{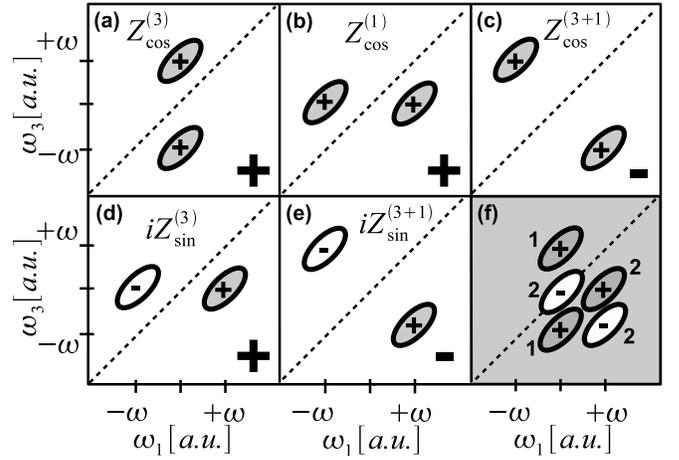}
\caption{\label{fig:shapeDemo}Demonstration of the construction of the 2D
line-shape $S_{R}^{cos}(\omega_{3},\omega_{1})$. Real parts of the
individual contributions are depicted in panels (a) to (e), together
with the signs of the particular contributions. The total line-shape
$S_{R}^{cos}(\omega_{3},\omega_{1})$ is depicted in panel (f). The
central negative feature originates from the unshifted line shape
$Z(\omega_{3},\omega_{1})$. The signs in the lower right corner in
panels (a) to (e) represent the signs with which the diagrams contribute
in Eq. (\ref{eq:ScosR}). The numbers in panel (f) denote the predicted
relative intensity of the features.}
\end{figure}

In this section, all calculations are performed assuming $T=300$ K and the bath correlation function in the form of an overdamped Brownian oscillator with reorganization energy $\lambda_{bath}=100$ cm$^{-1}$ and correlation time $\tau_{bath}=150$ fs. These numbers represent some characteristic values motivated by the molecular system studied in Paper I and Ref. \cite{Nemeth08a}. For the Brownian oscillator model, the long time line-shape function has a constant term $g_{const}=-\lambda \tau_{c}(2k_{B}T\tau_c/\hbar-i)$ and a linear term, which can be written as $\bar{g}(t)=-g_{const}t/\tau_c$. The high temperature limit $k_B T \gg \hbar/\tau_c$ is valid  for these parameters.

If the influence of vibrational modes converges to zero, i.e. for
$\lambda/\omega\rightarrow0$, the total spectrum is formed by the
sum of the non-shifted rephasing shape $Z$ and the non-shifted non-rephasing
shape $Y$. Qualitatively, these line-shapes are the same as those
presented in Fig. \ref{fig:shapeSlow}. The real rephasing part of
the spectrum has a characteristic orientation of the positive peak
along the diagonal line, while the real non-rephasing part is elongated
along the anti-diagonal line. Both line-shapes have negative {}``wings''
stretching perpendicularly to the axis of elongation of the particular
spectrum. If combined, they form a symmetric shape of the total response
with solvent induced broadening only. The dispersive (imaginary) part
of the solvent spectrum is also formed by the rephasing and non-rephasing
elementary 2D line-shapes which are characterized by a nodal line
parallel to the elongation axis of their real counterparts. Thus,
the rephasing and non-rephasing dispersive parts differ from each other by the orientation
of their nodal lines. The combined spectrum has a nodal line parallel
to the $\omega_{1}$-axis.

The 2D line-shapes $Z$ and $Y$ represent the basic forms from which
all spectral components can be constructed by means of a shift by
the vibrational frequency $\omega$. We demonstrate this on the contribution
$S_{R}^{cos}$ (cf. Eq. (\ref{eq:ScosR})). Fig. \ref{fig:shapeDemo}a-e
depicts the individual terms in Eq. (\ref{eq:ScosR}) according to
Eqs. (\ref{eq:Zeq1}), (\ref{eq:Zeq2}) and (\ref{eq:Zeq3}), and \ref{fig:shapeDemo}(f) presents the
total contribution $S_{R}^{cos}$. The oval shapes represent a single
contour of the real part of the rephasing line-shape $Z$,
thus showing their characteristic orientation and position in the
spectrum. The numbers Fig. \ref{fig:shapeDemo}f represent
the intensity of the corresponding spectral feature. Empty contours
represent negative peaks, whereas shaded contours indicate positive
peaks. From this schematic picture one can draw several conclusions.
First, if the shift $\omega$ is small and all contributions overlap
significantly, the positive part of the sum is elongated along the
diagonal, with asymmetric negative parts along the anti-diagonal line.
The negative features are stronger in the lower right corner of the
2D spectrum. Second, for vanishing $\omega$ the total contribution
tends to zero, because all terms cancel. Consequently, for the same
Huang-Rhys factor, the modulation by slower (lower energy) modes is
weaker than the corresponding modulation by faster (higher energy)
modes. Third, for $\omega$ larger than the width of the elementary
line-shapes $Z$, the rephasing $cos$ contribution has the strongest
positive feature on the right from the central frequency, a strong
negative feature in the lower right corner, while in the center the
contribution is negative. In the following sections we will demonstrate
this prediction by numerical evaluation of Eqs. (\ref{eq:StotSpectrum})
to (\ref{eq:ScosNR}). Similar intuitive pictures can be created also
for the non-rephasing spectra, using the pictorial representation
of the line-shape $Y$.

We will distinguish between fast (high energy) and slow (low energy)
vibrational modes relative to the inverse homogeneous spectral width
$\Delta_{solv}$ (FWHM) of the solvent contribution. If $\omega\ll\Delta_{solv}$,
i.e. vibrational crosspeaks in the 2D-FT spectrum or vibrational lines
in the absorption spectrum would be indistinguishable, we assume the
vibrational mode to be slow. The opposite case, $\omega\gg\Delta_{solv}$,
characterizes fast vibrational modes for which individual vibrational
features could be distinguished in optical spectra.

\begin{figure}[ht]
\includegraphics[clip,width=\columnwidth]{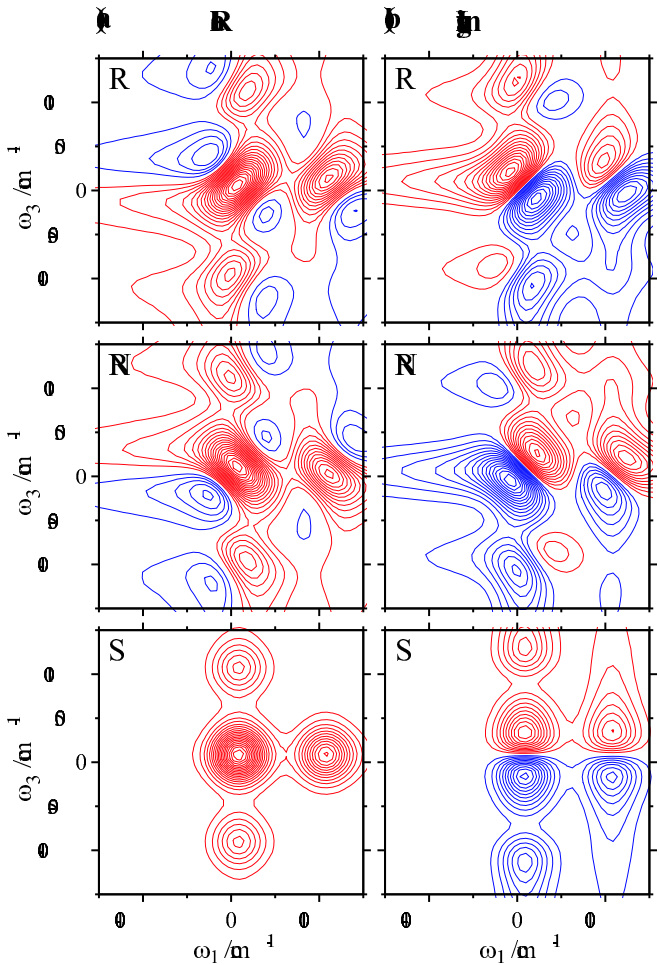}

\caption{\label{fig:shapeFast}The stationary contributions (from top down)
$S_{R}^{0}(\omega_{3},\omega_{1})$, $S_{NR}^{0}(\omega_{3},\omega_{1})$
and their sum $S^{0}(\omega_{3},\omega_{1})$ for the modulation by
a high energy vibrational mode of $\omega=1000$ cm$^{-1}$and $\lambda=300$
cm$^{-1}$. The figures on the left present the real and the figures
on the right the imaginary parts of the spectra, respectively.}
\end{figure}

Although the vibrational contribution is mostly dynamic, i.e. $t_{2}-$dependent,
there is a certain time-independent contribution to the total 2D spectrum
that originates from the vibrations. The stationary parts of the 2D
spectra for modulation by a low energy vibrational mode of $\omega=140$
cm$^{-1}$ and a high energy vibrational mode of $\omega=1000$ fs
are presented in Figs. \ref{fig:shapeSlow} and \ref{fig:shapeFast},
respectively. The reorganization energies have been set to $\lambda=80$
cm$^{-1}$ and $\lambda=300$ cm$^{-1}$, respectively. In case of
the low energy mode, the shift of the line-shape by $\omega$ does
not lead to any crosspeaks separated from the main diagonal peaks.
The observed line-shape is only slightly broadened and distorted from
the pure solvent line-shape by the vibrational contribution. In case
of the high energy mode, we can distinguish separate vibrational cross-
and diagonal peaks. In all cases we can identify the characteristic
direction of elongation of the 2D line-shapes. The imaginary parts
of all spectra also show characteristic orientations of the nodal
lines. We can conclude that, in the static part of the 2D spectrum
modulated by a high energy mode, the line-shapes are dictated by the
chromophore solvent interaction, while the positions of the peaks
are given by the vibrational frequency.

\subsection{Modulation of 2D spectra by a low energy mode}

The $t_{2}-$time evolution of the 2D spectra is mediated by the $\sin(\omega t_{2})$
and $\cos(\omega t_{2})$ functions which vary the magnitude of the
contributions in Eqs. (\ref{eq:SsinR}), (\ref{eq:ScosR}), (\ref{eq:SsinNR})
and (\ref{eq:ScosNR}). A closer look at Eqs. (\ref{eq:SsinR}) to
(\ref{eq:ScosNR}) reveals that the rephasing and non-rephasing contributions
have a very similar functional form. They differ mainly by the exchange
$Z\longleftrightarrow Y$. Apart from this exchange, comparing the
two $sin$ contributions, Eqs. (\ref{eq:SsinR}) and (\ref{eq:SsinNR}),
and the two $cos$ contributions, Eqs. (\ref{eq:ScosR}) and (\ref{eq:ScosNR}),
there is a flip of the sign in their first terms. This suggests that
the rephasing and non-rephasing modulations of the total 2D shape
differ not only in the orientation of their elongation, but also in
phase. For $\cos\omega t_{2}=1$, the total rephasing contribution
is at its maximum, while at the same time, the non-rephasing one is
at its minimum. Due to the third term in Eqs. (\ref{eq:SsinR}), (\ref{eq:ScosR}),
(\ref{eq:SsinNR}) and (\ref{eq:ScosNR}) this property is not exact.
However, by studying diagrams similar to Fig. \ref{fig:shapeDemo}
for the various contributions, we have concluded that a $cos$ non-rephasing
line-shape can be obtained from the rephasing one by turning the picture
by $\pi/2$ anti-clockwise (corresponding to $Z\longleftrightarrow Y$
exchange), mirroring it by the diagonal line, and changing the sign.

\begin{figure}[ht]
\includegraphics[width=\columnwidth]{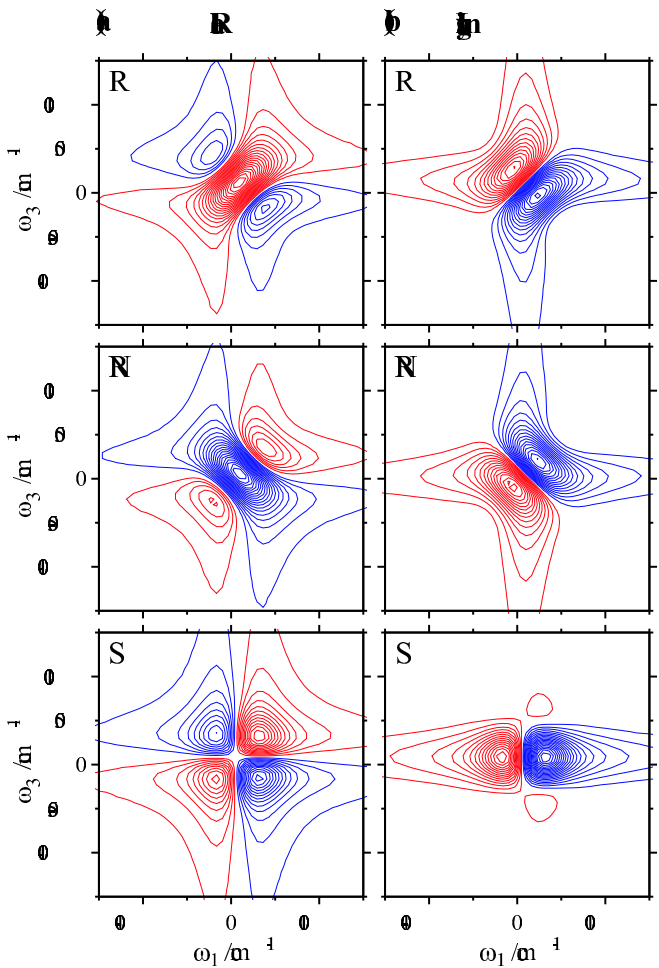}

\caption{\label{fig:cosOscSlow}The $cos$ contribution for the modulation
by a low energy vibrational mode of $\omega=140$ cm$^{-1}$. From
top down: the rephasing and non-rephasing parts, and the sum spectrum.
The figures on the left present the real and the figures on the right
the imaginary parts of the spectra, respectively.}

\end{figure}
\begin{figure}[ht]
\includegraphics[width=\columnwidth]{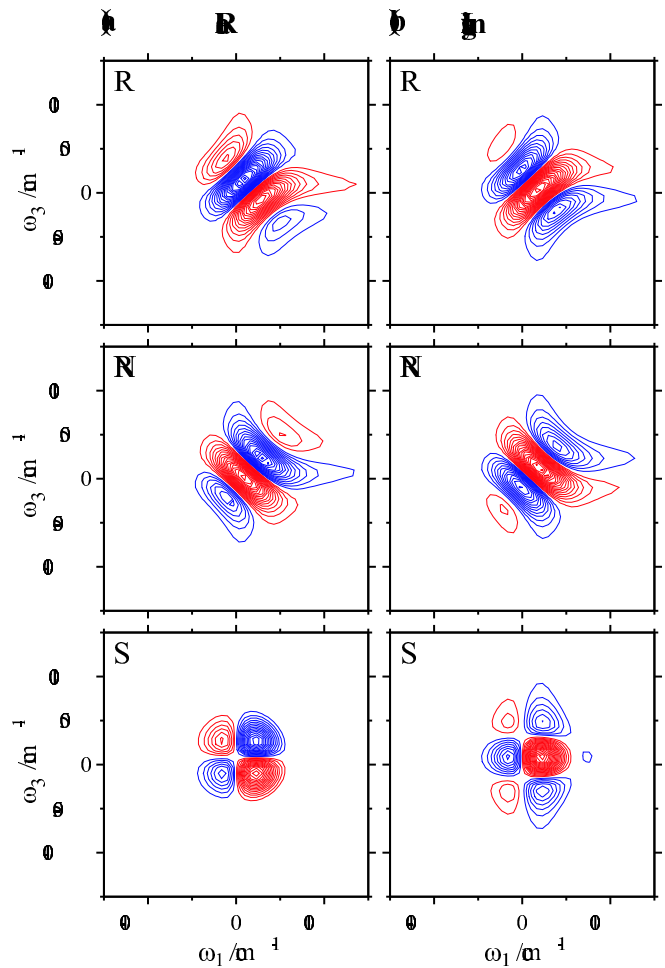}

\caption{\label{fig:sinOscSlow}The $sin$ contribution for the modulation
by a low energy vibrational mode of $\omega=140$ cm$^{-1}$. From
top down: the rephasing and non-rephasing parts, and the sum spectrum.
The figures on the left present the real and the figures on the right
the imaginary parts of the spectra, respectively.}

\end{figure}

\begin{figure}[ht]
\includegraphics[clip,width=\columnwidth]{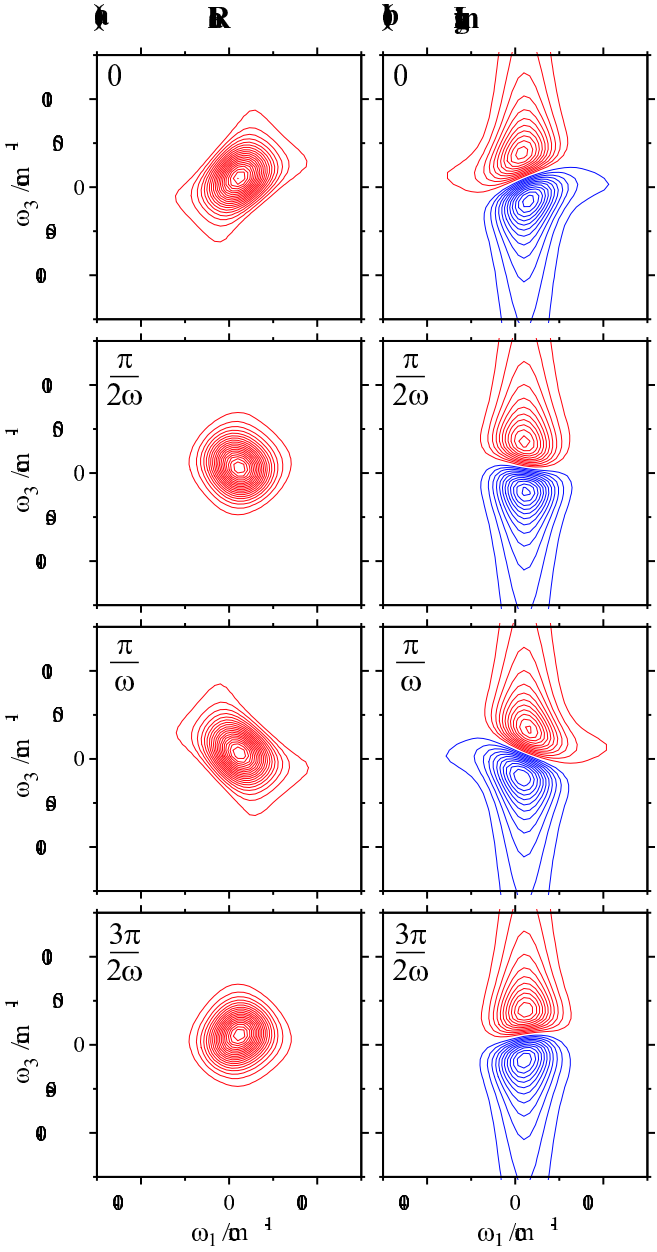}

\caption{\label{fig:slowMovie}The time evolution of the total 2D spectrum
of an electronic transition modulated by a low energy vibrational
mode of $\omega=140$ cm$^{-1}$. From top down: 2D spectra at times
$T=0,\mbox{\ensuremath{\frac{\pi}{2\omega}}, \mbox{\ensuremath{\frac{\pi}{\omega}}},}$
and $\frac{3\pi}{2\omega}$ fs. The figures on the left present the
real and the figures on the right the imaginary parts of the spectra,
respectively.}

\end{figure}

Figs. \ref{fig:cosOscSlow} and \ref{fig:sinOscSlow} present the
$cos$ and $sin$ oscillating parts of the total 2D spectrum for modulation
by a low energy vibrational mode. We can immediately notice that for
both the $sin$ and the $cos$ modulation the real rephasing and non-rephasing
contributions are approximately out of phase. For the set of parameters
used here, the maximum amplitude of the $cos$ contribution is approximately
twice the maximum amplitude of the $sin$ contribution, so that the
$cos$ oscillation is the most significant feature in the $t_{2}$-evolution
of the 2D spectrum. Neglecting therefore the $sin$ contribution for
a while, we can predict a mutual out of phase modulation of the rephasing
and non-rephasing parts of the 2D spectrum. This leads to the characteristic
oscillation of the diagonal and anti-diagonal width of the total 2D
spectrum (cf. Fig. \ref{fig:slowMovie}), which was demonstrated experimentally
in the preceding paper, Ref. \cite{Nemeth09a}. The nodal line rotation
in the dispersive part of the total 2D spectrum has a similar origin.
As the rephasing and non-rephasing parts of the dispersive contribution
have different orientation of their nodal lines, the total 2D spectrum
exhibits oscillations of the nodal line angle, due to the change of
relative amplitudes of rephasing and non-rephasing parts.

Fig. \ref{fig:slowMovie} demonstrates how the combined non-rephasing
and rephasing signal gives rise to the characteristic experimentally
observed modulation of the 2D line-shape. The results of our approximate
analysis in this section are well confirmed by the experimental 2D
spectra presented in Ref. \cite{Nemeth09a}. Quantitative agreement
with the experiment was achieved in Ref. \cite{Nemeth08a} by modeling
with the full expression for the corresponding response function,
Eqs. (\ref{eq:Rreph}) to (\ref{eq:R4}).

\subsection{Modulation of 2D spectra by a high energy mode}

\begin{figure}[ht]
\includegraphics[clip,width=\columnwidth]{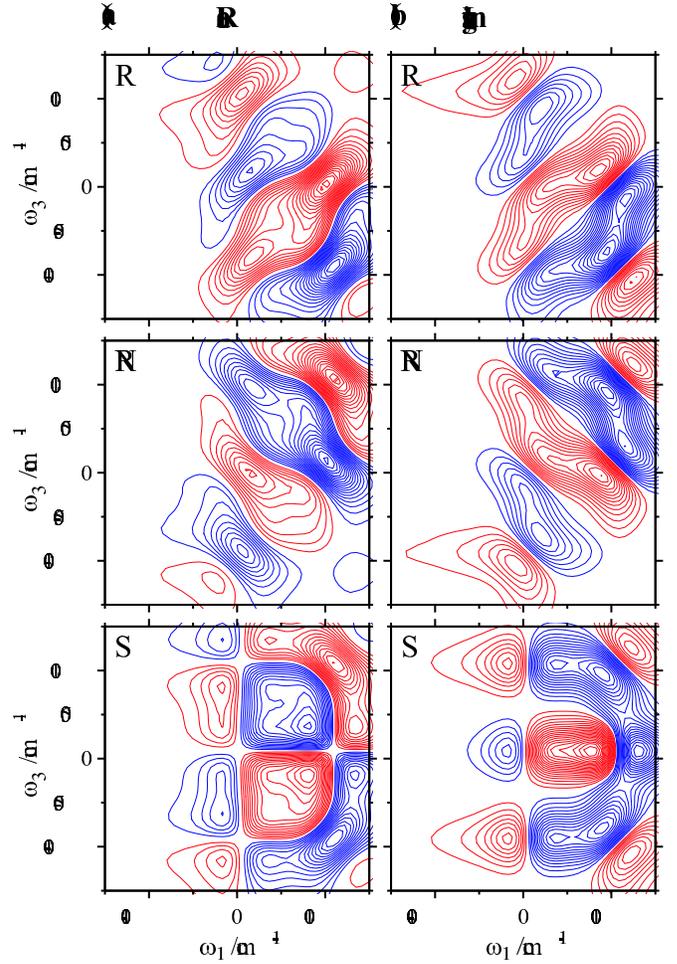}

\caption{\label{fig:cosOscFast}The $cos$ contribution for the modulation
by a high energy vibrational mode of $\omega=1000$ cm$^{-1}$. From
top down: the rephasing and non-rephasing parts, and the sum spectrum.
The figures on the left present the real and the figures on the right
the imaginary parts of the spectra, respectively.}

\end{figure}

\begin{figure}[ht]
\includegraphics[clip,width=\columnwidth]{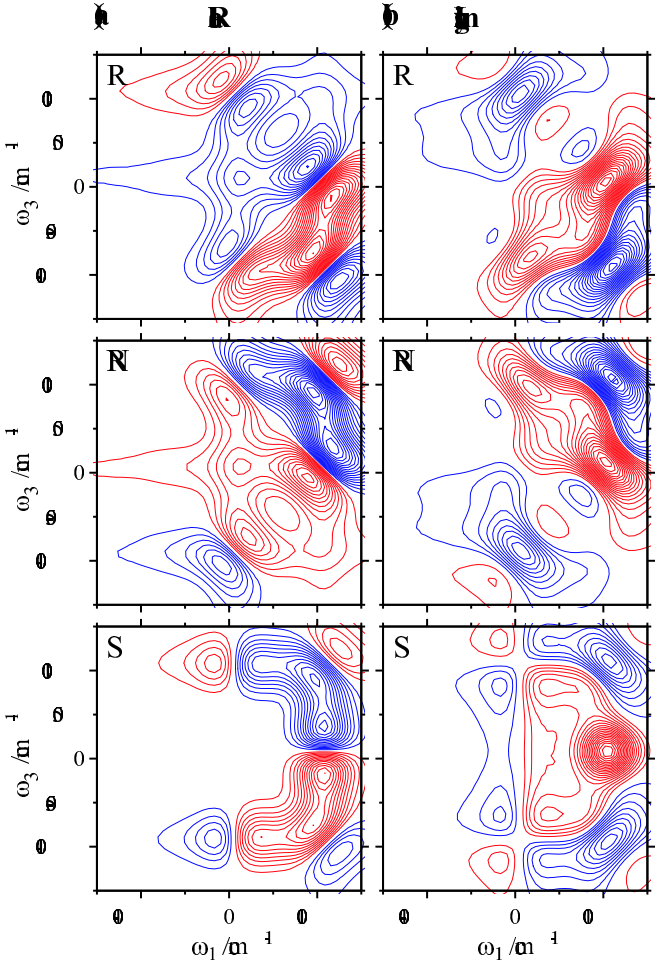}

\caption{\label{fig:sinOscFast}The $sin$ contribution for the modulation
by a high energy vibrational mode of $\omega=1000$ cm$^{-1}$. From
top down: the rephasing and non-rephasing parts, and the sum spectrum.
The figures on the left present the real and the figures on the right
the imaginary parts of the spectra, respectively.}

\end{figure}

Encouraged by the successful comparison of the theory presented here
with an experiment involving low energy vibrational modes, we can
extend the treatment to high energy modes. For fast vibrational modes,
where $\hbar\omega\gg2k_{B}T$, the validity of the expansion in Eqs.
(\ref{eq:R2_app}) and (\ref{eq:R1_app}) is limited only by the value
of the Huang-Rhys factor $\lambda/\omega$, because $\coth(\hbar\omega/2k_{B}T)\approx1$.
The composition of different contributions to the 2D spectrum in this
case, can be easily understood directly from considerations similar
to those which led to Fig. \ref{fig:shapeDemo}. The rephasing real
part of the $cos$ contribution shown in Fig. \ref{fig:cosOscFast}
corresponds very well to the depiction of Fig. \ref{fig:shapeDemo}f.
A negative peak is present in the center of the spectrum, while two
positive peaks are below the diagonal and one above the diagonal.
The stronger of the two positive peaks below the diagonal is positioned
at higher $\omega_{1}$ frequencies. All other contributions can be
constructed in a similar way, including those oscillating with $sin(\omega t_{2})$
(cf. Fig. \ref{fig:sinOscFast}).


\begin{figure}[ht]
 \includegraphics[clip,width=\columnwidth]{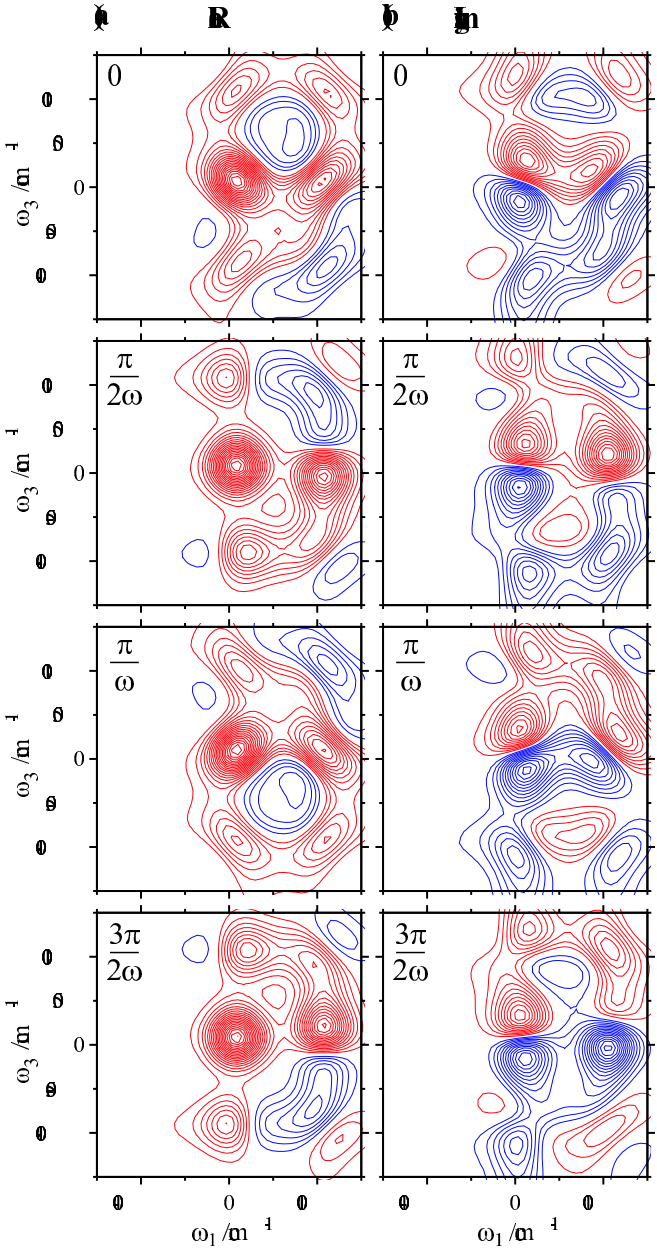}

\caption{\label{fig:fastMovie}The time evolution of the 2D spectrum of an
electronic transition modulated by a high energy vibrational mode
of $\omega=1000$ cm$^{-1}$. From top down: 2D spectra at times $T=0,\mbox{\ensuremath{\frac{\pi}{2\omega}}, \mbox{\ensuremath{\frac{\pi}{\omega}}},}$
and $\frac{3\pi}{2\omega}$ fs. The figures on the left present the
real and the figures on the right the imaginary parts of the spectra,
respectively.}

\end{figure}

Figure \ref{fig:fastMovie} presents the time evolution of the total
2D spectrum at different phases of the oscillation. An interesting
observation is that for high energy vibrational modes the diagonal
peaks oscillate with the opposite phases than in case of the low energy
mode modulation, and with the opposite phase with respect to some
of the cross-peaks. The shape of the $cos$ contributions for low
and high energy mode modulations (Fig. \ref{fig:cosOscSlow} vs. Fig.
\ref{fig:cosOscFast}), which are constructed from the same $Z$ and
$Y$ but with different amount of shift from the transition frequency,
again enables to understand the difference. Unlike the central diagonal
peak, the crosspeaks oscillate between positive and negative values.
The phase of these oscillations enables, for example, to clearly distinguish
the 2D spectra in times $t_{2}=\pi/2\omega$ and $t_{2}=3\pi/2\omega$
by the real part of the spectrum only, without referring to the
imaginary parts. For low energy modes, these two phases might be very
difficult to distinguish.

The presence of negative features in different parts of the 2D-FT
spectrum might influence the measured 2D-FT spectrum even if the
spectrum of the pulse is not able to cover vibrational crosspeaks
(see e.g. Ref. \cite{Nemeth09c}). However, sufficient information
for the characterization of wavepacket motion cannot be obtained from
such a measurement. If the experiment is performed with a laser pulse
of a bandwidth comparable with the width of the vibrational peaks
in the absorption or 2D spectrum (such as in Refs. \cite{Nemeth08a,Nemeth09a}
and \cite{Nemeth09c}) the frequency of the laser pulse relative
to the main diagonal peak determines the amount of influence the fast
mode has on the 2D spectrum. In particular, in Refs. \cite{Nemeth08a,Nemeth09a}
the central laser pulse frequency is lower than the position of the
main diagonal peak, and the 2D spectrum is thus measured away from
the fast vibrational features. Any vibrational modes with vibrational
frequency significantly faster than the width of the laser pulse can
thus be ignored from the evaluation, which was successfully done in
Ref. \cite{Nemeth08a}. On the other hand, if the laser pulse falls
into the spectral region between the diagonal peaks, as is the case
in Ref. \cite{Nemeth09c}, the 2D spectrum is distorted by the high
energy mode even if its frequency is larger that the width of the
laser pulse.

Application of ultra broadband pulses that would cover significantly
the whole region where a studied vibrational mode absorbs, promises
to enable a complete characterization of the photo-induced vibrational
wavepacket motion. Our results present reference spectra for an underdamped
wavepacket motion in a harmonic potential. Distortions from this reference
case can reveal changes in the shape of the wavepacket due to weak
relaxation, anharmonicity, and also differences between the ground
and excited state potential energy surfaces. From this point of view,
2D-FT experiments are of interest not only for molecules in solutions,
but also for gas phase molecules, were they can characterize fundamental
properties of vibrational DOF.

\section{Conclusions\label{sec:Conclusions}}

We have presented a qualitative theory of the time dependence of two-dimensional
line-shapes of a two-level electronic system weakly coupled to a vibrational
mode. We have studied the time evolution of the two-dimensional spectra
in case of low and high energy vibrational modes and showed that the
two-dimensional spectrum at $t_{2}$-times larger than the correlation
time of the solvent can be qualitatively predicted based on the purely
solvent two-dimensional line-shapes and knowledge of the vibrational
frequency and Huang-Rhys factor. The theory predicts oscillations
of the diagonal and anti-diagonal width of the two-dimensional spectrum
caused by the rephasing and non-rephasing signals being modulated
with opposite phase. In case of a low energy vibrational mode, the
signal is predominantly modulated by a cosine term, as confirmed in
the accompanying experimental paper. For high energy modes, where
the line-shape width is smaller than the vibrational frequency, our
theory predicts cross-peaks oscillating similarly to the low energy
mode case, but the diagonal peaks oscillating with a phase opposite
to the cross-peaks. Ultra broadband laser pulses would be necessary
to observe high energy modes of molecules in liquid phases. However,
in gas phase, a full characterization of vibrational motion based
on two-dimensional spectroscopy would be possible.

\vspace{1cm}
{\bf Acknowledgements.} T. M. acknowledges the kind support by the
Czech Science Foundation through grant GACR 202/07/P278 and by
the Ministry of Education, Youth, and Sports of the Czech
Republic through research plan MSM0021620835. This work was supported by
the Austrian Science Foundation (FWF) within the projects No.
P18233 and F016/18 \textit{Advanced Light Sources} (ADLIS). A.N.
and J.S. thank the Austrian Academy of Sciences for partial
financial support by the Doctoral Scholarship Programs
(DOCfFORTE and DOC).

\appendix
\section{Components of the vibrational response\label{sec:AppA}}

In this appendix, we present explicite formulae for the components of the vibrational response, Eq. (\ref{eq:Ffce}). The $t_{2}$-independent
part $K_{n}(t_{1},t_{3})$ of Eq. (\ref{eq:Ffce}) reads
\[K_{1}(t_{1},t_{3})=-i[\sin\omega t_{1}-\sin\omega t_{3}]\]
\begin{equation}
+\Xi(T)[\cos\omega t_{1}+\cos\omega t_{3}],\label{eq:K1}\end{equation}
\[K_{2}(t_{1,}t_{3})=i[\sin\omega t_{1}+\sin\omega t_{3}] \]
 \begin{equation}
+\Xi(T)[\cos\omega t_{1}+\cos\omega t_{3}],\label{eq:K2}\end{equation}
\[
K_{3}(t_{1},t_{3})=i[\sin\omega t_{1}-\sin\omega t_{3}]\]
 \begin{equation}
+\Xi(T)[\cos\omega t_{1}+\cos\omega t_{3}],\label{eq:K3}\end{equation}
 \[
 K_{4}(t_{1},t_{3})=-i[\sin\omega t_{1}+\sin\omega t_{3}]\]
 \begin{equation}
+\Xi(T)[\cos\omega t_{1}+\cos\omega t_{3}].\label{eq:K4}\end{equation}
The cosine oscillating part $H_{n}(t_{1},t_{3})\cos\omega t_{2}$ of Eq. (\ref{eq:Ffce})
can be written using\[
H_{1}(t_{1},t_{3})=\Xi(T)[1-\cos\omega t_{3}-\cos\omega t_{1}+\cos\omega(t_{1}+t_{3})]\]
 \begin{equation}
+i[\sin\omega t_{1}-\sin\omega t_{3}-\sin\omega(t_{1}+t_{3})],\label{eq:H1}\end{equation}
 \[
H_{2}(t_{1},t_{3})=-\Xi(T)[1-\cos\omega t_{3}-\cos\omega t_{1}+\cos\omega(t_{1}+t_{3})]\]
 \begin{equation}
+i[\sin\omega t_{1}-\sin\omega t_{3}-\sin\omega(t_{1}+t_{3})],\label{eq:H2}\end{equation}
 \[
H_{3}(t_{1},t_{3})=-\Xi(T)[1-\cos\omega t_{3}-\cos\omega t_{1}+\cos\omega(t_{1}+t_{3})]\]
 \begin{equation}
+i[\sin\omega t_{1}+\sin\omega t_{3}-\sin\omega(t_{1}+t_{3})],\label{eq:H3}\end{equation}
 \[
H_{4}(t_{1},t_{3})=\Xi(T)[1-\cos\omega t_{3}-\cos\omega t_{1}+\cos\omega(t_{1}+t_{3})]\]
 \begin{equation}
+i[\sin\omega t_{1}+\sin\omega t_{3}-\sin\omega(t_{1}+t_{3})].\label{eq:H4}\end{equation}
Finally, the sine oscillating part $G_{n}(t_{1},t_{3})\sin\omega t_{2}$
of Eq. (\ref{eq:Ffce}) has a form of\[
G_{1}(t_{1},t_{3})=\Xi(T)[\sin\omega t_{3}+\sin\omega t_{1}-\sin\omega(t_{1}+t_{3})]\]
 \begin{equation}
+i[1-\cos\omega t_{3}+\cos\omega t_{1}-\cos\omega(t_{1}+t_{3})],\label{eq:G1}\end{equation}
 \[
G_{2}(t_{1},t_{3})=-\Xi(T)[\sin\omega t_{3}+\sin\omega t_{1}-\sin\omega(t_{1}+t_{3})]\]
 \begin{equation}
+i[1-\cos\omega t_{3}+\cos\omega t_{1}-\cos\omega(t_{1}+t_{3})],\label{eq:G2}\end{equation}
 \[
G_{3}(t_{1},t_{3})=-\Xi(T)[\sin\omega t_{3}+\sin\omega t_{1}-\sin\omega(t_{1}+t_{3})]\]
 \begin{equation}
-i[1-\cos\omega t_{3}-\cos\omega t_{1}+\cos\omega(t_{1}+t_{3})],\label{eq:G3}\end{equation}
 \[
G_{4}(t_{1},t_{3})=\Xi(T)[\sin\omega t_{3}+\sin\omega t_{1}-\sin\omega(t_{1}+t_{3})]\]
 \begin{equation}
-i[1-\cos\omega t_{3}-\cos\omega t_{1}+\cos\omega(t_{1}+t_{3})].\label{eq:G4}\end{equation}

\section{2D line-shape components\label{sec:AppB}}
This appendix lists some intermediate definitions used to derive Eqs. (\ref{eq:S0R}) to (\ref{eq:ScosNR}) for the time dependent 2D line-shapes. Line-shape components complementing Eq. (\ref{eq:Ztcos}) read
 \[Z_{cos}^{(1)}(\omega_{3},\omega_{1})=FT_{2D}^{(-)}[e^{i(\omega_{eg}+\lambda)(t_{1}-t_{3})}\]
 \begin{equation}
\times \bar{R}_{2}(t_{3},t_{1})\cos\omega t_{1}](\omega_{3},\omega_{1}),\label{eq:Ztaucos}\end{equation}
 \[Z_{sin}^{(3)}(\omega_{3},\omega_{1})=FT_{2D}^{(-)}[e^{i(\omega_{eg}+\lambda)(t_{1}-t_{3})}\]
 \begin{equation}
\times\bar{R}_{2}(t_{3},t_{1})\sin\omega t_{3}](\omega_{3},\omega_{1}),\label{eq:Ztsin}\end{equation}
 \[Z_{sin}^{(1)}(\omega_{3},\omega_{1})=FT_{2D}^{(-)}[e^{i(\omega_{eg}+\lambda)(t_{1}-t_{3})}\]
 \begin{equation}
\times\bar{R}_{2}(t_{3},t_{1})\sin\omega t_{1}](\omega_{3},\omega_{1}),\label{eq:Ztausin}\end{equation}
 \[Z_{cos}^{(3+1)}(\omega_{3},\omega_{1})=FT_{2D}^{(-)}[e^{i(\omega_{eg}+\lambda)(t_{1}-t_{3})}\]
 \begin{equation}
\times\bar{R}_{2}(t_{3},t_{1})\cos\omega(t_{3}+t_{1})](\omega_{3},\omega_{1}),\label{eq:Ztautcos}\end{equation}
 \[Z_{sin}^{(3+1)}(\omega_{3},\omega_{1})=FT_{2D}^{(-)}[e^{i(\omega_{eg}+\lambda)(t_{1}-t_{3})}\]
 \begin{equation}
\times\bar{R}_{2}(t_{3},t_{1})\sin\omega(t_{3}+t_{1})](\omega_{3},\omega_{1}).\label{eq:Ztautsin}\end{equation}

Lineshape components complementing Eq. (\ref{eq:Zeq1}) read
 \begin{equation}
Z_{cos}^{(1)}(\Omega_{3},\Omega_{1})=\frac{1}{2}\left(Z^{(0+)}+Z^{(0-)}\right),\;  
\end{equation}
\begin{equation}
Z_{sin}^{(1)}(\Omega_{3},\Omega_{1})=\frac{i}{2}\left(Z^{(0+)}-Z^{(0-)}\right),\label{eq:Zeq2}\end{equation}
 \begin{equation}
Z_{cos}^{(3+1)}(\Omega_{3},\Omega_{1})=\frac{1}{2}\left(Z^{(+-)}+Z^{(-+)}\right),\; \end{equation}
 \begin{equation}
Z_{sin}^{(3+1)}(\Omega_{3},\Omega_{1})=\frac{i}{2}\left(Z^{(-+)}-Z^{(+-)}\right),\label{eq:Zeq3}\end{equation}
 \begin{equation}
Y_{cos}^{(3)}(\Omega_{3},\Omega_{1})=\frac{1}{2}\left(Y^{(+0)}+Y^{(-0)}\right),\; \end{equation}
 \begin{equation}
Y_{sin}^{(3)}(\Omega_{3},\Omega_{1})=\frac{i}{2}\left(Y^{(-0)}-Y^{(+0)}\right),\label{eq:Zeq4}\end{equation}
 \begin{equation}
Y_{cos}^{(1)}(\Omega_{3},\Omega_{1})=\frac{1}{2}\left(Y^{(0+)}+Y^{(0-)}\right),\; \end{equation}
 \begin{equation}
Y_{sin}^{(1)}(\Omega_{3},\Omega_{1})=\frac{i}{2}\left(Y^{(0-)}-Y^{(0+)}\right),\label{eq:Zeq5}\end{equation}
 \begin{equation}
Y_{cos}^{(3+1)}(\Omega_{3},\Omega_{1})=\frac{1}{2}\left(Y^{(++)}+Y^{(--)}\right),\; \end{equation}
 \begin{equation}
Y_{sin}^{(3+1)}(\Omega_{3},\Omega_{1})=\frac{i}{2}\left(Y^{(--)}-Y^{(++)}\right).\label{eq:Zeq6}\end{equation}

\end{document}